\documentclass[11pt]{article}
\usepackage[nomove]{overcite}
\usepackage{amsmath}
\usepackage{amssymb}
\usepackage{epsfig}
\usepackage[small]{caption2}
\usepackage{xspace}
\usepackage{indentfirst}
\newcommand{\beq}{\begin{equation}}
\newcommand{\eeq}{\end{equation}}
\newcommand{\chiab}{\ensuremath{\chi_{\text{AB}}}\xspace}
\newcommand{\chiastarb}{\ensuremath{\chi_{\text{A*B}}}\xspace}
\newcommand{\chiaastar}{\ensuremath{\chi_{\text{AA*}}}\xspace}
\newcommand{\xI}{\ensuremath{x_{\text{I}}}\xspace}
\newcommand{\xII}{\ensuremath{x_{\text{II}}}\xspace}
\newcommand{\dotxI}{\ensuremath{\dot{x}_{\text{I}}}\xspace}
\newcommand{\dotxII}{\ensuremath{\dot{x}_{\text{II}}}\xspace}
\newcommand{\Ds}{\ensuremath{D_{\text{s}}}\xspace}
\newcommand{\Df}{\ensuremath{D_{\text{f}}}\xspace}
\newcommand{\dd}{\ensuremath{\text{d}}\xspace}
\newcommand{\muA}{\ensuremath{\mu_{\text{A}}}\xspace}
\newcommand{\phic}{\ensuremath{\phi_{\text{c}}}\xspace}
\DeclareMathOperator{\erf}{erf}
\begin{document}
\title{\sffamily \bfseries Interfacial layering in\\ a three-component polymer system}
\author{A.\ Aradian$^{1,2}$\footnote{A.Aradian@ed.ac.uk},
F.\ Saulnier$^1$\footnote{Florent.Saulnier@college-de-france.fr},
E.\ Rapha\"{e}l$^1$\footnote{Elie.Raphael@college-de-france.fr}
and P.-G.\ de
Gennes$^1$\footnote{PGG@curie.fr}\\
$^1${\footnotesize Physique de la Mati\`{e}re Condens\'{e}e,
C.N.R.S.\ UMR 7125 \& F\'{e}d\'{e}ration de Recherche MSC (FR 2438),}\\
{\footnotesize Coll\`{e}ge de France, 11
place Marcelin Berthelot, 75231 Paris Cedex 05, France}\\
$^2$ {\footnotesize University of Edinburgh, School of Physics,}\\
{\footnotesize King's Buildings JCMB, Edinburgh EH9 3JZ, United
Kingdom}} \maketitle
\begin{abstract}
We study theoretically the temporal evolution and the spatial
structure of the interface between two polymer melts involving
\emph{three} different species (A, A* and B). The first melt is
composed of two different polymer species A and A* which are
fairly indifferent to one another (Flory parameter
$\chiaastar\simeq 0$). The second melt is made of a pure polymer B
which is strongly attracted to species A ($\chiab<0$) but strongly
repelled by species A* ($\chiastarb>0$). We then show that, due to
these contradictory tendencies, interesting properties arise
during the evolution of the interface after the melts are put into
contact: as diffusion proceeds, the interface structures into
several adjacent ``compartments'', or layers, of differing
chemical compositions, and in addition, the central mixing layer
grows in a very asymmetric fashion. Such unusual behaviour might
lead to interesting mechanical properties, and demonstrates on a
specific case the potential richness of multi-component polymer
interfaces (as compared to conventional two-component interfaces)
for various applications.
\end{abstract}
\section{Introduction and motivation}
The phenomena taking place at the interfaces and surfaces of
polymeric systems are obviously of high importance to many
practical or industrial situations. As a result, they have been
the focus of extensive study both experimentally and theoretically
in the recent decades. Much progress has been made in the
understanding of these phenomena as well as in the development of
techniques to characterize them (see e.g. the general
references~\citen{JonesRichardsPolymerInterfaces,WoolPolymerInterfaces,
SanchezFitzpatrickPolymerInterfaces,RichardsPeacePolymerSurfacesInterfaces}).
\subsection{Interfaces between polymer melts}
One situation which has received much attention is that of two
pieces of molten polymer put into (good) contact at a certain
initial time: how does then the interface between these melts
evolve, and how does the mixing between the melts occur, if any?
or, more precisely, what is the final, equilibrium state reached
by the system (i.e., given enough time, does it mix fully or only
in a restricted region?), and with what kind of concentration
profile? What is the dynamics leading to that final state, over
which typical timescales? etc. It has been
found,~\cite{JonesRichardsPolymerInterfaces,WoolPolymerInterfaces}
due to the specific physics of polymer macromolecules, that their
mixing is generally quite different from what is known in more
conventional diffusive systems, like molecular gases.

Once the two polymer pieces are put into contact, the subsequent
evolution will obviously generally depend on the various
physico-chemical properties of the materials facing each other
(nature of microscopic interactions between monomers, chain
length, topological structure, \ldots). It has been recognized
however that, at least from a conceptual standpoint, the two most
crucial parameters determining the fate of the system are the sign
and magnitude of the product $\chi N$, where $\chi$ is the Flory
parameter,~\cite{PGGScalingConcepts} related to the microscopic
interactions between components, and $N$ is the chain length. We
will remind of the different possibilities for $\chi N$ in
section~2.
\subsection{Multi-component interfaces}
In order to facilitate the identification and modelling of the
fundamental processes at work within polymer/polymer interfaces,
past studies have for the most part, if not exclusively, focused
on interfaces involving \emph{two} species (for instance, one on
each side of the interface) and sometimes only one species
(interfaces between two identical melts). Certainly, the span of
real situations is much wider, with a vast range of
\emph{multi-component interfaces}, and one is thus entitled to ask
what is the structure and dynamics of an interface between two
polymer \emph{mixtures} involving more than two components.

As exemplified on the case studied in this article, such
multi-component interfaces may have an interest of their own, as a
means (for example) of obtaining \emph{structured interfaces at a
sub-micron scale}: when two polymer mixtures are put into contact,
and provided the contrast in relative miscibilities of the
different components is adequate, a ``multiple'' interface may
form; that is to say the interface will partition into several
adjacent layers, each of different chemical composition. A
possible application of this could be the formation of multilayer
polymer films. The mechanical properties presented by the
resulting assembly might also prove interesting.

Multi-species interfaces can also appear in the course of an
interfacial reaction between two polymers: when the polymer pieces
are put into contact, a reaction may start at the interface that
delivers a (polymer) product whose miscibilities with each of the
initial reactants are different; this three-species system is
susceptible to lead to a multiple interface, whose dynamics is
then coupled to that of the chemical reaction.

However, on purely combinatorial grounds, the diversity of all
conceivable multi-component interfaces makes any general approach
likely to be hopeless; thus delineating and focusing on a certain
number of limiting situations may be useful. This is indeed the
purpose of the present article: we will be considering one such
``selected case'' for a three-component interface, and will show
that it exhibits a peculiar spatial structure and evolution
through time, which we will fully analyze.

The article is organized as follows: in section~2, we start with a
reminder on the standard theoretical results concerning
two-component interfaces, as these will be used as basic tools for
the rest of the paper. In section~3, we present the system that
will be studied, and try to qualitatively explain the structure of
the multiple interface which appears. In section~4, we write the
equations governing the system and solve them, thus finding the
lengthscales and concentration profiles which characterize the
dynamics of our system (note that, if interested only in results,
our reader can skip directly to section 4.3). Section~5 closes the
article with some concluding remarks.
\section{A short overview of two-component interfaces}
\label{reminder}
We start here by a general overview of the standard theories
describing two-species interfaces. By no means do we aim here at
an exhaustive survey of the theoretical or experimental literature
on the subject: we simply quickly recall the most basic
theoretical elements on which our present work builds.

Let us denote the two polymer species we consider as A and B. We
will cover interfaces occurring between an A-melt facing a B-melt,
as well as between two A-B blends facing each other but of
differing A/B proportions. We will also mainly put the emphasis on
the effect of varying the \emph{Flory parameter} $\chiab$ between
A and B, with the assumption that both A and B have the
\emph{same} chain length $N$. Situations where $N_\text{A}$ and
$N_\text{B}$ differ (e.g. a melt of long chains facing a melt of
short chains), which are still subject to some debates, will
merely be hinted at.

It is a remarkable point that the kinetics of formation of
interfaces between chemically different polymers is controlled by
thermodynamic as well as kinetic factors. To take into account the
driving or slowing role played by energetic parameters in
diffusion processes, it is much convenient to use Flory-Huggins'
derivation of the free energy (see eq.~\ref{FloryFreeEnergy} for
the expression of the free energy $f$ in this model, and also
note~\citen{flory}): thinking in terms of a lattice model, the
dimensionless \emph{Flory parameter} $\chiab$ characterizes the
enthalpy of mixing at the molecular level, by comparing
interactions between neighboring polymer segments of the same
species (\emph{i.e.}, association energy $\epsilon_\text{AA}$ for
two neighboring A segments, $\epsilon_\text{BB}$ for two B
segments) and interactions between different species in contact
($\epsilon_\text{AB}$ for an A next to a B). If $z$ denotes the
coordination number in the lattice, we define
\begin{equation}
\chiab=
\frac{z(2\epsilon_\text{AB}-\epsilon_\text{AA}-\epsilon_\text{BB})}{2
k T}
\end{equation}
In words, $\chiab$ is the energy change, in units of the thermal
energy $k T$, when a segment of A is taken from an environment of
pure A and swapped with a segment of B from an environment of pure
B. The Flory parameter $\chiab$ can be positive or, much more
rarely, negative. If the only interactions existing between A and
B are van der Waals forces, $\chiab$ shall be positive. Negative
values of $\chiab$ do appear in polymer couples displaying
specific chemical interactions, such as hydrogen bonds. The three
following subsections briefly discuss the different possible cases
($\chiab>0$, $\chiab=0$, $\chiab<0$) and their physical
consequences.
\subsection{Immiscible components: \boldmath{$\chiab>0$}}
As stated above, in the absence of specific interactions or
structural similarities, the Flory parameter is positive and
usually ranges from $10^{-3}$ to $10^{-1}$. This is by far the
most usual situation for polymer pairs. For polymer pairs
differing only by isotopic substitution (e.g., when deuterium is
substituted to hydrogen in the monomer structure), very small,
positive values as low as $\chiab \simeq 10^{-4}$ can be
found.\cite{bateswignall}

In this case of positive $\chiab$, the polymer pairs are
\emph{immiscible} for all but the lowest molecular weights: if
$N=N_\text{A}=N_\text{B}$ is the length of both polymers A and B,
it can be shown that, for a Flory parameter $\chiab$ greater than
a critical value $\chi_\text{C}=2/N \ll 1$, phase separation
occurs and we end up at equilibrium with macroscopic regions of
pure A and pure B, separated by a sharp interfacial layer: no
macroscopic mixing occurs.

Let us precise the nature of this interface between the two
coexisting phases by a simple scaling argument.\cite{helfand} The
characteristic width $w$ of the interface can be estimated by a
balance between the chain entropy (of order $kT$ per chain), which
tends to widen the interface, and the unfavorable enthalpy of
mixing (of order \chiab per segment), which tends to narrow the
interface: thus the typical length $N_0$ of a loop of an A-chain
penetrating a B-rich region is given by $N_0\chiab k T \sim k T$,
that is, $N_0 \sim 1/ \chiab$. Because the loop has the form of a
random statistical walk, this contour length corresponds to a
(straight) penetration distance $w \sim a \sqrt{N_\text{max}}$
(where $a$ is the monomer size). We finally find that the
characteristic width of the interface is
\begin{equation}
 w \sim \frac{a}{\sqrt{\chiab}}
\label{interfacewidth}
\end{equation}

This simple analysis is in accordance with the results obtained
from a more rigorous description based on a square-gradient model
of the interfacial free energy.\cite{helfand} Note that the
scaling analysis leading to this result is valid provided that
$\chiab N \gg 1$: equation~\ref{interfacewidth} then predicts the
mixing region to be much smaller than the polymer coil size, and
thus, at the macroscopic scale, the interface is to be considered
as very sharp one, with a steep profile. This property of
interfaces between immiscible melts will be used thereafter in our
solution (cf. Sec.~\ref{qualitativeapproach}).
\subsection{Entropy-driven mixing: \boldmath{$\chiab=0$}}
We now consider the case of a zero Flory parameter. Rigorously
speaking, this situation arises when the two polymers put into
contact are identical (same chemical structure). But it is also a
useful approximation to describe mixing in systems with a very
small \chiab, be it positive or negative (typically when
$|\chiab|\ll 1/N$).

Starting from the initial situation where two polymer pieces are
put into contact, the final, equilibrium state reached by the
system is a homogeneous one (the initial interface has totally
disappeared) with complete mixing of the polymer components. As
there is no enthalpy gain associated with mixing species
($\chiab=0$), the evolution of the system is driven solely by
gains in the translational entropy of the polymer chains.

The dynamics towards this final state can be complex, especially
for entangled polymers where several regimes of non-Fickean
diffusion successively appear. In these entangled polymers, the
first stages of the interdiffusion process between melts of equal
molecular weight have been described first within a scaling
approach in Ref.\citen{degennessoudurepolymamorph}, and then
detailed by different
authors:\cite{pragertirell,kimwool,adolftirellprager,zhangwool}
the chains are initially segregated on each side of the contact
plane, with discontinuous concentration profiles abruptly dropping
from unity to zero at the separation. Then, as the chains start
interpenetrating to form a mixed layer, and because of  the
specific reptational dynamics of the polymer chains, this initial
discontinuity survives, with a progressively resorbing amplitude,
until disappearing when the reptation time $T_\text{rep}$ is
reached. The interface is then said to have ``healed'' completely.
This feature stands in obvious contrast with conventional mixing
where initial discontinuities are immediately smoothed out by the
diffusion process.

For times greater than the reptation time ($t>T_\text{rep}$), the
interdiffusion dynamics becomes purely Fickean, and is described
by the classical diffusion equation
\begin{equation}
\label{entropicdiffusion} \dot\phi = \Ds \nabla^2\phi
\end{equation}
The diffusion coefficient $D_\text{s}$, known as the
``self-diffusion'' coefficient, is independent of concentration,
and for long, entangled chains ($N$ greater than the entanglement
threshold $N_\text{e}$), is equal to
\begin{equation}
\label{entropiccoeffdiff}
D_\text{s}=\Lambda_0\frac{N_\text{e}}{N^2}kT
\end{equation}
where $\Lambda_0$ is the monomeric mobility.

The use of eq~\ref{entropicdiffusion} is valid only for facing
melts with chains A and B of equal length (symmetric interfaces
with $N_\text{A}=N_\text{B}$). The case of \emph{asymmetric}
polymer junctions, between chemically identical melts of different
molecular weights (typically a melt of short chains in contact
with a melt of much longer chains) proves much more subtle. The
evolution of such a system also displays several temporal regimes
and can essentially be understood as
follows:\cite{PGGfrancoise,brocharddegennes} the long and less
mobile chains behave like a gel which is progressively penetrated
and swollen by the smaller species; but as the gel of long chains
swells, it effectively drags the smaller chains within it. The
global diffusion profile is thus rather complex and results from
the microscopic diffusion of the small chains relative to the
matrix of longer chains, combined with the global motion of this
matrix. This global flow of the matrix, known as a
``bulk-flow'',\cite{CrankDiffusion} is essential for a correct
description of the system and will reappear in the next section
devoted to enthalpy-driven species.
\subsection{Enthalpy-driven mixing: \boldmath{$\chiab<0$}}
\label{chiabnegatif}
Let us now consider the case of mixing between species driven by a
gain in enthalpy, i.e., when $\chiab<0$. A few dozens of A/B pairs
with $\chiab<0$ have been reported, like PS/PVME
(polystyrene/poly-vinylmethylether)~\cite{shibayama} or PVC/PMMA
(poly-vinylchloride)/poly-methylmethacrylate). Within the
temperature range where $\chiab$ remains negative, such polymer
couples are fully miscible for all molecular weights.

The dynamics of ``enthalpy-driven'' mixing at interfaces between
polymers has been the focus of many studies in the past 20 years,
and its theoretical description has led to some controversy in the
case where the system is \emph{asymmetric}, e.g., when the
mobility of the A and B chains are different (due to different
chain lengths).

Based upon Onsager's formalism of linear irreversible
thermodynamics,\cite{onsager,PGG} two main approaches were
proposed.

The so-called ``slow-mode'' theory propounded by Brochard and
co-workers \cite{BJL,BJL2}, assumes that the flux of species A is
the opposite of the flux of B, i.e., $J_\text{A}=-J_\text{B}$. In
cases where the system is asymmetric, this hypothesis of
cancellation of fluxes leads to a mutual diffusion coefficient
dominated by the mobility of the \emph{slowest} species---hence
the name of the theory.

On the other hand, Kramer and co-workers\cite{kramer} explicitly
introduce a third species in their model, which play the role of
vacancies in the system and move about in order to allow the
motion of A and B chains. Then the fluxes of the A and B
components need not be equal anymore, and their difference
$J_A-J_B$ is compensated by a net flux of vacancies $J_V$ across
the cell.\cite{sillescu} It is furthermore postulated that the
chemical potential of the vacancies is constant throughout the
system, and this, contrarily to the theory of Brochard et al.,
leads to a ``fast-mode'' diffusion coefficient, i.e., dominated by
the \emph{fastest} species in asymmetric systems.

In the recent years, several authors have tried to bridge the gap
between these ``slow'' and ``fast'' theories~; see for instance
refs.~\citen{jabbari,AKN} and references therein.

Experimental data seem to validate the fast-mode
theory.\cite{klein,gilmore,jordanball,seggern,meier} Nonetheless,
to this day, the origin and physical interpretation of the vacancy
flux used in the model of Kramer et al. or in more recent
approaches remains somewhat obscure. Therefore, it is conceptually
most interesting to note that, in an amended version of her
model,\cite{francoise} Brochard showed that fast-mode predictions
could be retrieved \emph{without} resorting to any flux of
vacancies.\cite{fastmode} Instead, Brochard's corrected approach
shows that, in addition to the individual diffusion fluxes of A
and B, there exists a global, \emph{collective motion} involving
\emph{both} A and B; when this is taken into account, fast-mode
results are found without the need of vacancies. The occurrence of
such collective ``bulk-flows'' (as they are called), superimposed
to pure diffusion, is indeed known as a common feature of
asymmetric diffusive systems.\cite{CrankDiffusion} In the amended
model of Brochard, each chain of polymer reptates and diffuses
within a matrix composed of all the other chains; physically, a
bulk-flow occurs because this matrix \emph{itself} is set into
motion, due to the difference of mobility between the two
diffusing species (in other words, polymer chains reptate in tubes
which are themselves moving).

To describe the interdiffusion process of these two different
polymers, one usually employs the so-called ``mutual diffusion''
coefficient $D_\text{M}$, which relates the \emph{total} flux
$J_i$ of species $i$ ($i=$A or B) to the gradient of its volume
fraction $\phi_i$: $J_i=-D_\text{M} \nabla \phi_i$. Let us denote
$\phi \equiv \phi_A=1-\phi_B$ the volume fraction of species A. In
all the theoretical models mentioned above, the mutual diffusion
coefficient can then be cast into the generic form:
\begin{equation}
\label{coeffgenform} D_\text{M}= \Lambda[\phi] \cdot \phi(1-\phi)
\cdot \frac{\mbox{d}^2 f}{\mbox{d} \phi^2}
\end{equation}
where $\Lambda[\phi]$ is a positive function of $\phi$
(homogeneous to a mobility), which differs from model to model,
and $f$ is the classical Flory-Huggins free energy per
site,\cite{flory} given by:
\begin{equation}
\frac{f}{k T}= \frac{\phi}{N_\text{A}}
\log{\phi}+\frac{1-\phi}{N_\text{B}} \log{(1-\phi)}+ \chiab \phi
(1-\phi) \label{FloryFreeEnergy}
\end{equation}

In the rest of this article, we will only be concerned with the
\emph{symmetric} case, i.e., with polymer components of the same
chain length ($N_\text{A}=N_\text{B} \equiv N$) and the same
monomeric mobility $\Lambda_0$. It should be pointed out that, in
this case, the conceptual subtleties mentioned above do not
intervene, and interdiffusion models do agree on the same
expression of the diffusion coefficient. For long, entangled
chains ($N>N_\text{e}$, the regime of interest to us), and using
the same notations as in eq.~\ref{entropiccoeffdiff}, one has
simply $\Lambda=\Lambda_0 N_\text{e}/N$. From
eq~\ref{coeffgenform}, the mutual diffusion coefficient then
reads\cite{BJL,BJL2,kramer}
\begin{equation}
\label{coefficiententangled} D_\text{M}= \Lambda_0
\frac{N_\text{e}}{N}kT\cdot \phi(1-\phi)\cdot \left[ \frac{1}{\phi
(1-\phi)N} - 2 \chiab\right]
\end{equation}
A useful limit in practice is $|\chiab|N \gg 1$: as long as $\phi$
and $1-\phi$ are larger than the (small) value $\phi_\text{c}=1/(2
|\chiab|N)$, $D_\text{M}$ can be approximated to
\begin{equation}
\label{simplecoeff} D_\text{M}= \Df
\phi(1-\phi)\quad\text{with}\quad \Df \equiv 2 |\chiab| N \Ds
\end{equation}
(where $D_\text{s}$ is the entropic, self-diffusion coefficient
defined in eq~\ref{entropiccoeffdiff}).

Equation~\ref{simplecoeff} shows that the diffusion coefficient in
the enthalpy-driven mixing of polymers is strongly
concentration-dependent. In particular, the vanishing of
$D_\text{M}$ for $\phi=0$ and $\phi=1$ leads to very unusual
diffusion profiles when either of the polymer pieces brought in
contact is initially pure in one of the components:\cite{BJL} When
only one side is pure, the mixing width on that side is
\emph{finite}, with a sharp edge; if both the initial polymer
pieces are pure, the width of the mixing region is finite on both
sides, and, moreover, the diffusion profile within it is a
\emph{straight line}.

(Of course, as stated above, the simplified form~\ref{simplecoeff}
of $D_\text{M}$ does not hold when $\phi\to 0$ or $\phi\to 1$, and
one has in fact to return to the full
expression~\ref{coefficiententangled} for which $D_\text{M}$
\emph{does not} exactly vanish\ldots Thus, strictly speaking, the
width of the mixing region is never exactly finite, and enthalpic
diffusion profiles near $\phi=0$ and $\phi=1$ cross over to
conventional, infinite ``diffusion tails''. But this crossover
occurs for $\phi$ or $1-\phi$ of order $\phi_\text{c}=1/(2
|\chiab|N)$, and in the limit where $|\chiab| N \gg 1$,
$\phi_\text{c}$ is so small that the diffusion tails contain a
completely negligible mass of material.)

With this, we conclude our overview on the different aspects of
the theory of two-component interdiffusion, and come back to the
problem of multi-component interfaces.
\section{Description of the three-component system and qualitative approach}
In this section, we present the specific multi-component system
that we chose to study, and give some qualitative insight into the
structure and dynamics that are expected to emerge. The next
section will deal with these issues in a more rigorous way.

As explained in the Introduction, the realm of multi-species
system and their interfaces is a very vast one: our aim here is to
explore one ``model case'' among these possibilities, involving
three species, and where some interesting features do appear.
\subsection{Description of the system and initial configuration}
\label{initialsystem}
We now present the system that we will be concerned with in the
rest of this article. The initial ($t=0$) situation is depicted in
Figure~1: the system consists of two pieces of molten polymer
which have just been put into contact. The surface of contact (the
initial interface) between the pieces is assumed to be planar, and
is located at position $x=0$ (the $x$-axis is drawn perpendicular
to the surface of contact). It is assumed that the problem is
invariant in the two remaining spatial directions (parallel to the
interface).
    \begin{figure}
    \centering
    \includegraphics*[scale=.8, clip=true, bb=4.3cm 10.7cm 16.5cm
    16.5cm]{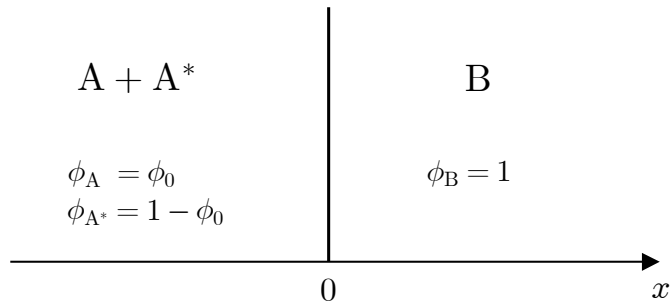}
    \caption{Initial situation of the three-species system.
    The left-hand side is a blend of two polymers,
    A and A*, with initial volume fractions $\phi_\text{A}=\phi_0$
    and $\phi_\text{A*}=1-\phi_0$. The right-hand side contains only
    one species, called B, with initial volume fraction $\phi_\text{B}=1$.}
    \end{figure}

The system contains \emph{three} polymer components which are
initially distributed as follows: on the left-hand side, we have a
blend of two polymers, A and A*; on the right-hand side, we have a
pure polymer, i.e., only one species which is called B. The
important step (which is really defining our system) is how to
specify the nature of the pairwise affinities between these three
components. Our choice is the following: we assume that A and B
are attracted to each other (through some specific interaction),
and are thus miscible in all proportions; on the contrary, A* and
B strongly repel each other and thus have a tendency to form fully
segregated phases; and, finally, A and A* are fairly indifferent
to each other (i.e., neither attracted or repelled, or not much).
These features correspond to the following set of Flory
parameters:
    \begin{alignat}{4}
    \label{defchiab}
    &\chiab &< 0, &\qquad\qquad &\text{with} &\quad& |\chiab|N &\gg 1\\
    \label{defchiastarb}
    &\chiastarb&>0, &&\text{with} &&\chiastarb N &\gg 1\\
    \label{defchiaastar}
    &\chiaastar &\simeq 0, &&\text{with} &&|\chiaastar| N &\ll 1
    \end{alignat}
As we are mainly interested in the effect of the contrast in
miscibility between the polymers, and not in chain length effects,
it has been assumed above that all polymers have the same
polymerization index:
    \begin{equation}
    N_\text{A}=N_\text{A*}=N_\text{B}=N.
    \end{equation}

A three-component system with such features as specified in
eqs~\ref{defchiab}--\ref{defchiaastar} would be quite difficult to
realize experimentally, as the above requirements seem at first
rather contradictory; they would certainly be achieved only
through a fine tuning of the chemistry of the different polymers
involved. We discuss some practical possibilities in the
concluding remarks of the article (Section~5).

Our choice of system is mainly guided here by the fact that it is
one case that leads to the most interesting interfacial structure
and dynamics. From the theoretical point of view, this interfacial
problem is also original in the sense that, as will be seen
shortly, it leads to simultaneous, coupled, entropy and
enthalpy-driven diffusion processes.
\subsection{Qualitative approach to the multiple interface structure}
\label{qualitativeapproach}

In this section, we present qualitative arguments explaining why,
from the initial situation depicted in Figure~1, the interface
between the two melts develops a ``multiple'' structure, i.e.
displays several layers or regions of different nature.

Quickly after the onset of contact between the two melts, the
interface takes on a configuration as depicted on Figure~2.
    \begin{figure}
    \centering
    \includegraphics*[scale=.8, clip=true, bb=4.2cm 9.7cm 16.5cm
    18.5cm]{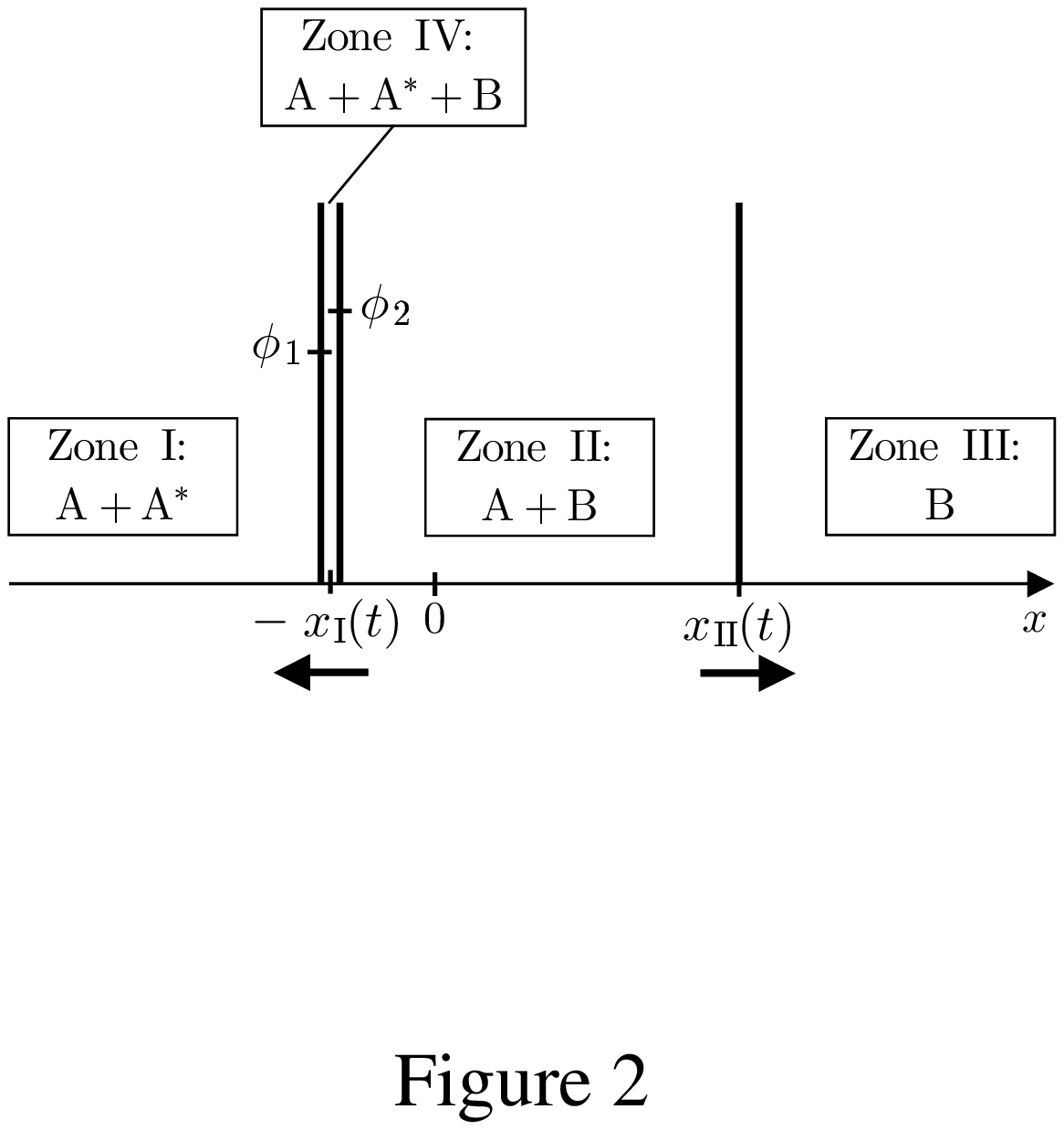}
    \caption{Multiple structure of the interface, with several regions, for $t>0$. The arrows
    indicate the direction of motion of the regions' boundaries,
    $x=-\xI(t)$  and $x=\xII(t)$. The quantities $\phi_1$ and $\phi_2$
    indicate the values of the volume fraction in A at the left and right
    boundaries of region~IV.}
    \end{figure}
As can be seen, the interface is divided in four main zones,
labelled I--IV, and the situation can indeed be intuitively
understood as explained now. (We do not consider here the
transient evolution which takes place in the first instants after
contact and leads into the configuration of Fig.~2. The interested
reader is referred to Appendix~A for a more detailed discussion of
this transient.)

After the onset of contact, species A and B (which are attracted
one to another) start to mix together, thus forming a growing
mixing layer (region~II) around the initial position ($x=0$) of
the surface of contact between the melts.

On the other hand, A* is repelled by B, and thus, as region~II
broadens and the A+B mixture spreads, A* must recede towards the
left. Thus region~I, which we define as the domain of coexistence
of A and A*, is progressively shrinking, with its right boundary
moving towards the left.

At the junction between region~I (containing A and A*) and
region~II (containing A and B), there must be a region where
\emph{all} three species coexist (region~IV). However, we expect
this region~IV to be of very small extension as compared to the
others, since we know from section~2.1 that strongly immiscible
polymers put into contact, like here A* and B, tend to form a very
sharp interface with a steep profile; hence, in the remaining of
this article, we will neglect the thickness of region~IV, and we
will not consider any detailed concentration profile within it. It
must be emphasized however that region~IV, albeit tiny, plays an
important role in the dynamics of the system, since all A chains
which want to migrate from region~I to II must cross region~IV
(this will be discussed again in the next section).

Finally, there is a last region, III, where B is alone: this comes
from the fact that, as explained in Sec.~\ref{chiabnegatif}, when
one of the polymer pieces in contact is initially pure (here,
$\phi_\text{B}=1$ at $t=0$ for the piece on the right),
enthalpy-driven interdiffusion generates profiles with mixing only
over a finite width; thus, there is on the right side of region~II
a (time-dependent) location where the volume fraction of A reaches
zero, and this is what defines the beginning of region~III. As the
mixing region~II broadens, the boundary of region~III shifts to
the right.

We also introduce in Figure~2 some notations that we will be using
intensively in the next sections: $x=-\xI(t)$ gives the position
of region~IV, and, in our limit where region~IV is taken
infinitesimally thin, plus or minus exponents ($x=-\xI^\pm$) will
respectively refer to the right and left boundaries of region~IV;
$x=\xII(t)$ gives the position of the boundary between regions~II
and III; and $\phi_1(t)$ and $\phi_2(t)$ respectively denote the
volume fractions of A just on the left and right of $-\xI$, i.e.,
$\phi_1=\phi_\text{A}|_{x=-\xI^-}$ and
$\phi_2=\phi_\text{A}|_{x=-\xI^+}$.

After having presented an intuitive description of the interfacial
structure and dynamics in our system, we will devote the next
section to the establishment and solution of the equations
governing them.
\section{Governing equations and results}
In this section, we first establish the various equations
governing the system (Sec.~4.1), and then solve them numerically
(Sec.~4.2). The physical results of our computations are gathered
and commented on in Sec.~4.3.
\subsection{Governing equations}
\label{governing}
From now on, $\phi_\text{A}$, the volume fraction of A, will be
simply denoted by $\phi$, i.e., we will use
$\phi(x,t)\equiv\phi_\text{A}(x,t)$.

Let us start by enumerating our unknowns: these are the locations
of the different regions' boundaries ($x=-\xI$ and $x=\xII$), the
values of the volume fraction in A at $x=-\xI^{\pm}$ ($\phi_1$ and
$\phi_2$), and the time-dependent profile of $\phi(x,t)$ within
regions~I and II (by definition, $\phi=0$ within region~III). We
will thus need a set of six independent equations to determine the
solution of the system.

It should be noted that, from the knowledge of $\phi$, it is
straightforward to retrieve the volume fractions of the other
components of the system (using the total volume fraction
condition $\phi+\phi_\text{A*}+\phi_\text{B}=1$): in region~I,
where only A and A* are present, one has $\phi_\text{A*}=1-\phi$
and $\phi_\text{B}=0$; in region~II, $\phi_\text{B}=1-\phi$ and
$\phi_\text{A*}=0$; finally, in region~III, $\phi_\text{B}=1$ and
$\phi=\phi_\text{A}=0$.

We now establish the six independent equations which govern the
dynamics of our system.
\subsubsection*{Diffusion equations}
Our first equations simply describe the diffusive motion of
polymer~A. We must use two different equations, depending on the
region considered within the multiple interface.

In region~I, A chains diffuse inside an A+A* mixture, with
$\chiaastar \simeq 0$ (eq~\ref{defchiaastar}); we are thus in a
case of entropy-driven diffusion which is accounted for by
equation~\ref{entropicdiffusion},
    \begin{equation}
    \label{diffeqI}
    \dot\phi - \Ds\nabla^2\phi=0 \qquad\text{(region~I)}
    \end{equation}
and the initial and boundary conditions are as follows:
    \begin{equation}
    \label{conditionsdiffeqI}
    \phi(t=0)=\phi_0; \qquad \phi(x\to-\infty)=\phi_0,\quad
    \phi(x=-\xI^-)=\phi_1
    \end{equation}
In eq~\ref{diffeqI}, the entropic diffusion coefficient \Ds is a
constant parameter (see eq~\ref{entropiccoeffdiff}).

In region~II, A chains diffuse into an A+B mixture, with $\chiab <
0$ and $|\chiab|N \gg 1$ (eq~\ref{defchiab}); we are there in the
presence of an enthalpy-driven diffusion, for which the diffusion
coefficient becomes concentration-dependent (see Sec.~2.3).
Recalling eq~\ref{simplecoeff}, we have
    \begin{equation}
    \Df=2|\chiab|N\Ds \qquad (\Df\gg\Ds)
    \end{equation}
and the diffusion equation in region~II can be written as
    \begin{equation}
    \label{diffeqII}
    \dot\phi - \Df\nabla.\bigl(\phi(1-\phi)\nabla\phi\bigr)=0 \qquad\text{(region~II)}
    \end{equation}
with the initial and boundary conditions
    \begin{equation}
    \label{conditionsdiffeqII}
    \phi(t=0)=0; \qquad \phi(x=-\xI^+)=\phi_2,\quad
    \phi(x\to+\infty)=0
    \end{equation}
(see note\cite{noteboundary}).
\subsubsection*{Expansion of the A-B mixing region on the right}
Apart from the two diffusion equations~\ref{diffeqI} and
\ref{diffeqII}, a third equation can be written, which governs the
expansion rate of the mixing region~II (A-B mixture) on the right,
that is to say, the motion of the boundary $\xII(t)$. By
definition, $x=\xII(t)$ denotes, for any time $t$, the moving
point where the concentration profile $\phi$ reaches zero (and
beyond which only polymer B is present). As shown in the
Appendix~B to this article, this very definition implies that the
\emph{velocity} $\dotxII$ of this point has a simple relation to
the gradient of $\phi$ on its left as follows:
    \begin{equation}
    \label{equationxIIpoint}
    \dotxII=-\Df\nabla\phi|_{x=\xII^-}
    \end{equation}
\subsubsection*{Conservation of the A and B species}
We must naturally also take into account equations that ensure the
conservation of the different polymer components in the system.

It is shown in Appendix~B that the conservation equation for
species A leads to the following relation between the velocity
$\dotxI$ of the interface of regions~I and~II, and the volume
fraction $\phi_1$ of A and the gradient $\nabla\phi|_{x=-\xI^-}$
on the left side of it:
    \beq
    \label{conservationAlocale}
    \dotxI=-\frac{1}{1-\phi_1}\cdot\Ds\nabla\phi|_{x=-\xI^-}
    \eeq

The conservation of the B species, on the other hand, gives a
relation between the same velocity $\dotxI$, and the volume
fraction and gradient on the \emph{right} side of this interface:
    \beq
    \label{conservationBlocale}
    \dotxI=-\phi_2\cdot\Df\nabla\phi|_{x=-\xI^+}
    \eeq

Note that the conservation of the third species, A*, follows
automatically from the conservation of the two others, through the
relation between volume fractions
$\phi+\phi_\text{A*}+\phi_\text{B}=1$, and thus does not bring us
a new equation.
\subsubsection*{The role of region~IV}
At this point, we have written down five independent relations; as
there are six unknowns, there must be a last physical constraint
determining the dynamics of the system. We have so far totally
ignored region~IV (as defined on Figure~2), where the three
species A, A* and B coexist. As stated earlier, this is because
this region is expected to be of very limited extension, and, as
such, has rightfully been discarded, e.g., from the A and B
conservation equations discussed above. However, despite being
negligible in mass, region~IV must clearly play an essential role
regarding the dynamics, since all the A material that diffuses
from region~I to region~II has to transit through it.

Describing precisely the interdependent diffusive process at work
within the three-body region~IV---solving the ``inner problem'',
in the terms of boundary-layer theory---would undoubtedly
represent a complex task and require a study of its own. We rather
propose to bypass this difficulty, as we in fact only need very
limited information for our purpose: we are only interested in the
effect of region~IV on the outer dynamics, not in internal
details.

We make the simple assumption that region~IV opposes no particular
resistance to the flow of polymer A passing through it: region~IV
is ``permeable'' to A as other regions in the system are, or in
other words, the mobility of species A in region~IV is not
significantly different from the mobility in region~I or~II. This
assumption on mobility appears natural if one considers that
although region~IV differs from other regions by significant
changes in composition (it is the only region where all three
species overlap), it does not drastically differ in microscopic
structure; hence, the reptational dynamics, which control
mobility, should not change significatively within region~IV as
compared to other regions. (The situation would completely be
different if region~IV was, for instance, made of a densely
crosslinked network which would hinder the passage of polymer
chains; one would then rather work under the opposite, ``high
resistivity'' assumption).

From this permeability assumption, we can deduce a relation on the
chemical potential of species A on each side of region~IV: a
chemical potential gradient must be present to drive the diffusion
of A through region~IV, but because this region is both of ``low
resistivity'' \emph{and} because it is very thin (as argued
earlier in the text), the overall drop of chemical potential over
region~IV will be minute as compared to the potential drop over
other (larger) regions, and can therefore be completely neglected.

Denoting $ \muA(-\xI^-)$ the chemical potential of species A just
on the left of region~IV, and $\muA(-\xI^+)$ the potential on the
right,\cite{notepotentiel} we thus have
    \beq
    \label{egalitepotentiel}
    \muA(-\xI^-)=\muA(-\xI^+)
    \eeq

As shown in Appendix~B, once the expressions of the chemical
potentials are written out, this equality between potentials
provides our last governing equation in the form of a relation
between the volume fractions $\phi_1$ and $\phi_2$ at the borders
of region~IV:
    \beq
    \label{equationphi1phi2}
    \phi_1=\phi_2\,\exp\bigl[-|\chiab|N(1-\phi_2)^2\bigr]
    \eeq

We are now in possession of six independent equations which govern
the dynamics in our system, and can proceed to solve them.
\subsection{Technical solution}
\label{technicalities}
We now present the procedure followed to solve the set of
governing equations derived in the previous section
(eqs~\ref{diffeqI}, \ref{diffeqII}, \ref{equationxIIpoint},
\ref{conservationBlocale}, \ref{conservationAlocale},
\ref{equationphi1phi2}). We will here focus on the technical
aspects of the solution, while, for the sake of readability, we
have grouped together the presentation of the results and their
physical content in the next section (Sec.\ref{results}).

The major difficulty attached to the interdiffusion problem as
defined by this set of equations is that it involves two
\emph{moving boundaries}, $x=-\xI(t)$ and $x=\xII(t)$: the
solution of the diffusion equations in the system requires the
application of boundary conditions defined at the moving
boundaries, but the motion of the boundaries themselves are
directly determined by the diffusive solutions. This
self-consistent nature of moving boundary problems make them
generally much less straightforward to solve than a conventional
diffusion problem; we will here follow the solution scheme
described in Crank.~\cite{CrankDiffusion}

Our problem falls into the so-called ``Class A''
category,~\cite{CrankDiffusion} characterized by the fact that the
motion of the moving boundaries is uniquely due to the transfer of
diffusing substances across them. (This is indeed the meaning of
the conservation equations~\ref{conservationBintegree1} and
\ref{conservationAintegree} discussed in Appendix~B). It is then
possible to show that, in a generic fashion, the solution will
have the following properties:~\cite{CrankDiffusion}

(i)~The concentration of the diffusing species at the moving
boundaries is constant, i.e., $\phi_1$ and $\phi_2$ are
\emph{independent of time}.

(ii)~The motion of the moving boundaries is \emph{diffusive}, i.e.
$\xI(t)\sim(\Ds t)^{1/2}$ and $\xII(t)\sim(\Df t)^{1/2}$.

We build the solution to our problem with the help of these useful
results. In accordance with property~(ii) above, we introduce the
new quantities $a$ and $b$ defined by:
    \begin{eqnarray}
    \label{defa}
    \xI(t)&=& 2a\,(\Ds t)^{1/2}\\
    \label{defb}
    \xII(t)&=&2b\,(\Df t)^{1/2}
    \end{eqnarray}
The numerical value of $a$ and $b$ will have to be determined as a
result of our solution. We also define
    \beq
    \label{defepsilon}
    \epsilon=\frac{\Ds}{\Df}=\frac{1}{2|\chiab|N} \quad \ll 1
    \eeq
which is a small parameter (see eqs~\ref{simplecoeff}
and~\ref{defchiab}).

In the limit where $\epsilon \ll 1$,
equation~\ref{equationphi1phi2} simplifies considerably:
    \beq
    \label{phi1egalezero}
    \phi_1=\phi_2\,\exp\Bigl(-\frac{(1-\phi_2)^2}{2\epsilon}\Bigr)
    \underset{\epsilon \ll 1}{\longrightarrow} 0
    \eeq
Thus, as long as $(1-\phi_2)\gg\epsilon^{1/2}$ (i.e., as long as
$\phi_2$ is not too close to unity), we can as well consider that
$\phi_1=0$ for the rest of our solution to a very good
approximation.

We can then immediately solve the diffusion
equation~\ref{diffeqI}, valid in region~I, along with the boundary
conditions~\ref{conditionsdiffeqI}. The solution can as usual be
looked for as a function $\phi(x,t)=f(u)$ of the reduced variable
$u=x/2\sqrt{\Ds t}$, and from eqs~\ref{defa} and
\ref{phi1egalezero}, we see that the boundary condition on the
moving boundary simply becomes $\phi(-\xI^-,t)=f(-a)=0$. Solving
the equation on $f$, we finally find the following concentration
profile in region~I, which involves the error function $\erf
u=(2/\sqrt{\pi})\int_0^u e^{-u^2}\,\dd u$, classical for
semi-infinite media:
    \beq
    \label{solutionregionI}
    \phi(x,t)= \phi_0 - \frac{\phi_0}{1-\erf a}
    \left[1+ \erf\left(\frac{x}{2\sqrt{\Ds t}}\right)\right]
    \qquad \text{(region I)}
    \eeq

We next turn to the determination of $a$, and thus to the
determination of the position of the moving boundary at $x=-\xI$.
Equation~\ref{conservationAlocale} will serve this purpose: with
$\phi_1=0$, it simplifies to $\dotxI=-\Ds\nabla\phi|_{x=-\xI^-}$.
Then, with the help of eq~\ref{defa}, which yields
$\dotxI=a(\Ds/t)^{1/2}$, and of eq~\ref{solutionregionI} which
yields an expression of $\nabla\phi|_{x=-\xI^-}$,
eq~\ref{conservationAlocale} finally brings the relation
    \beq
    \label{equationsura}
    \sqrt{\pi} \,a \,e^{a^2} \,(1-\erf a)=\phi_0
    \eeq
This equation on $a$ can be solved numerically to give the value
of $a$ as a function of the fundamental parameter $\phi_0$ (which
describes the initial state of the system). The result of the
numerical solution is shown in Figure \ref{aandb}.

The next step is to solve the diffusion equation~\ref{diffeqII}
under the conditions~\ref{conditionsdiffeqII}, in order to find
the concentration profile in region~II. As equation~\ref{diffeqII}
is a nonlinear equation without analytical solution (to the best
of our knowledge), we will resort to numerical integration. We
here again look for a solution $\phi(x,t)=g(v)$ of the (new)
reduced variable $v=x/2\sqrt{\Df t}$. The resulting differential
equation on $g$ turns out as:
    \beq
    \label{equationsurg}
    \frac{\dd}{\dd v}\Bigl(g(1-g)\frac{\dd g}{\dd v}\Bigr)
    =-2v\,\frac{\dd g}{\dd v}
    \eeq
and the boundary conditions of eq~\ref{conditionsdiffeqII}
translate to
    \beq
    \label{conditionssurg}
    \text{(i)}\quad \phi|_{x=-\xI^+}=g|_{v=-a\sqrt\epsilon}=\phi_2,
    \qquad\text{(ii)}\quad \phi|_{x\to+\infty}=g|_{v\to+\infty}=0
    \eeq
It has to be noted that these boundary conditions, as such, are
not sufficient to directly proceed to numerical integration,
because the quantity $\phi_2$ appearing there is \emph{also} an
unknown. We thus need a supplementary condition, which will be
provided by the conservation equation of the B species
(eq~\ref{conservationBlocale}): using the definition of $a$
(eq~\ref{defa}) and the definition of the reduced variable $v$,
eq~\ref{conservationBlocale} can be rewritten as
    \beq
    \label{conditionsupplementaire}
    \frac{\dd g}{\dd v}\Bigr|_{v=-a\sqrt\epsilon}
    =-\frac{2a\sqrt\epsilon}{\phi_2}
    \eeq

We have now enough equations to compute $g(v)$ and the associated
concentration profile in region~II. We solve the set of
equations~\ref{equationsurg}--\ref{conditionsupplementaire} by a
trial-and-error procedure: we choose an arbitrary value for the
unknown $\phi_2$; based on the corresponding initial conditions
provided by eqs~\ref{conditionssurg}-(i) and
\ref{conditionsupplementaire}, a numerical integration of
eq~\ref{equationsurg} is performed; the obtained solution $g$ is
then evaluated at infinity, and its value compared to zero
(eq~\ref{conditionssurg}-(ii)); if different, the whole numerical
procedure is resumed with a new chosen value for $\phi_2$, until
the correct value of $\phi_2$, satisfying
eq~\ref{conditionssurg}-(ii), is found. At the end of this
process, we eventually obtain the actual profile of concentration
$\phi(x,t)=g(v)$ in region~II, along with the value of $\phi_2$ in
our system. (See further down for results.)

We are at this point in possession of the profiles $\phi(x,t)$ in
regions~I, II and III, as well as of the numerical values of $a$,
$\phi_1$, and $\phi_2$. To complete our solution, we still need to
find the value of $b$, which is related to the position of the
border between region~II and III. This is easily performed
numerically: by transposing the equation of motion for \xII
established earlier (eq~\ref{equationxIIpoint}) in terms of the
self-similar function $g(v)$, and applying the definition of $b$
(eq~\ref{defb}), one obtains the relation
    \beq
    -2b=\frac{\dd g}{\dd v}\Bigr|_{v=b}
    \eeq
i.e., $b$ is a quantity such that the slope of $g$ at $v=b$ is
equal to twice the value of $b$ (with changed sign). The search
for such a $b$ can be carried on numerically with the function $g$
calculated previously: the graph of $b$ as a function of $\phi_0$
can be found on Figure \ref{aandb}.
\subsection{Presentation of the results}
\label{results}
We now present and comment on the results of the model, as
obtained through the procedure described in the previous section.

As seen in eqs~\ref{defa} and~\ref{defb}, the different
``compartments'' or layers composing the multiple interface are
separated by diffusive fronts: $x=-\xI(t)=- 2 a (\Ds t)^{1/2}$ and
$x=\xII(t)=2 b (\Df t)^{1/2}$, with $a$ and $b$ are numerical
parameters (see below for their values). The evolution in time of
the concentration profiles of the different components is thus
self-similar, and for this reason, it is useful to present them
here in terms of the reduced variable $u=x/2 (\Ds t)^{1/2}$.

We remind our reader that there are two main physical parameters
to our solution: $\phi_0$, which represents the content in A of
the initial A+A* blend, and the small parameter
$\epsilon=1/(2|\chiab|N)\ll 1$, which is related to the strength
of the attraction between A and B chains and the polymer chain
length.

As these parameters were varied over a wide range of values
($\phi_0$ between 0.1 and 0.9, $\epsilon$ between $10^{-5}$ and
0.1), it was observed that, remarkably enough, the (self-similar)
diffusion profiles obtained retain the same aspect throughout; the
main effect of varying the parameters is simply to \emph{rescale}
one or the other of the characteristic dimensions of the
self-similar solution (i.e., stretch or shrink it in some region).
\begin{figure}
\centering
\includegraphics*[scale=.7, clip=true, bb=1.6cm 8cm 18.5cm
17cm]{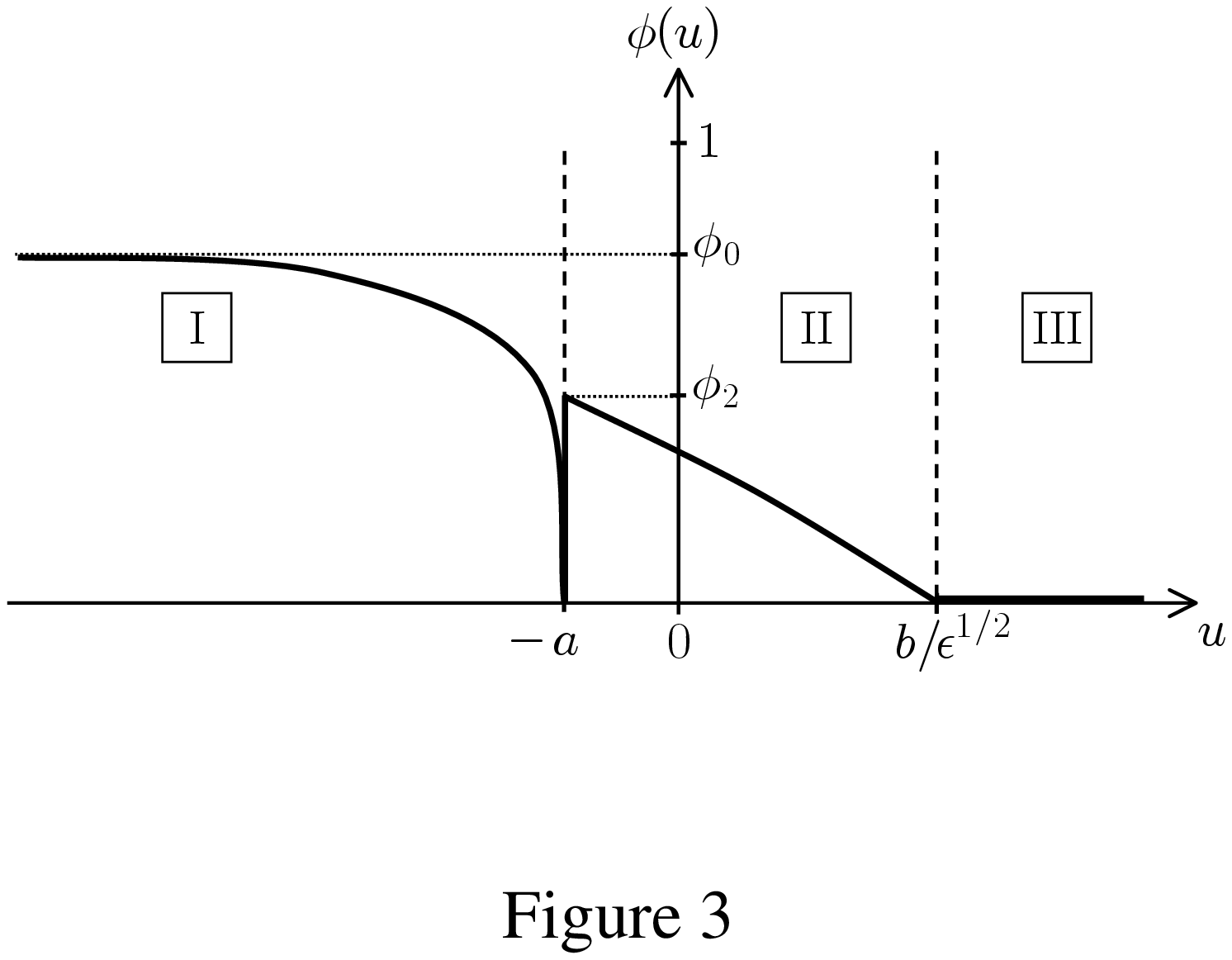} \caption{Sketch of the generic self-similar
diffusion profile for species A: the volume fraction $\phi$ has
been drawn versus the reduced variable $u=x/2 \sqrt{\Ds t}$. The
volume fractions used in the text, $\phi_0$ and $\phi_2$, are also
represented. In domain~I (\emph{i.e.} for $u<-a$), the
concentration profile is an error function; in domain~II
(\emph{i.e.} for $-a<u<b/\epsilon^{1/2}$), the profile is
quasi-linear. In region~III, A is absent.} \label{profilgenerique}
\end{figure}

This generic self-similar diffusion profile for the A species is
sketched in Figure~\ref{profilgenerique} in terms of the reduced
variable $u$. The fronts $x=-\xI(t)$ and $x=\xII(t)$ are
respectively located at $u=-a$ and $u=b/\sqrt{\epsilon}$. Starting
from the left in the Figure, the profile in region~I is given by
an ``error function'', typical of diffusion problems, and drops to
zero at the approach of region~II (since we proved that
$\phi_1\simeq0$ in most practical cases). At the boundary between
regions~I and II (we neglect the thickness of the intermediate
region~IV), there is a discontinuous concentration jump from
$\phi=0$ to $\phi=\phi_2$, and then the diffusion profile
decreases smoothly until reaching zero again at the border with
region~III. It should be emphasized as a very interesting feature
of the solution that the amplitude $\phi_2$ of this concentration
jump remains constant in time and \emph{does not} die away during
the diffusion process.

Another interesting feature is the depletion near $u=-a$ of
species~A ($\phi\to 0$). The depletion is a direct reflection of
the fact that it is energetically much more favorable for A to be
in the A+B blend of region~II ($\chiab<0$) than in the A+A* blend
of region~I ($\chiaastar\simeq0$); therefore, as time passes,
region~II continuously sucks A-chains out of region~I, creating in
the latter a depleted zone which grows with time (as $\sqrt{\Ds
t}$).

We also note that, from an empirical point of view, the profile in
region~II does not have a very significant curvature; if needed,
we may thus approximately regard it as a straight line.

\begin{figure}
\centering
\includegraphics*[scale=.65, clip=true, bb=3.2cm 6.2cm 18.5cm
17.5cm]{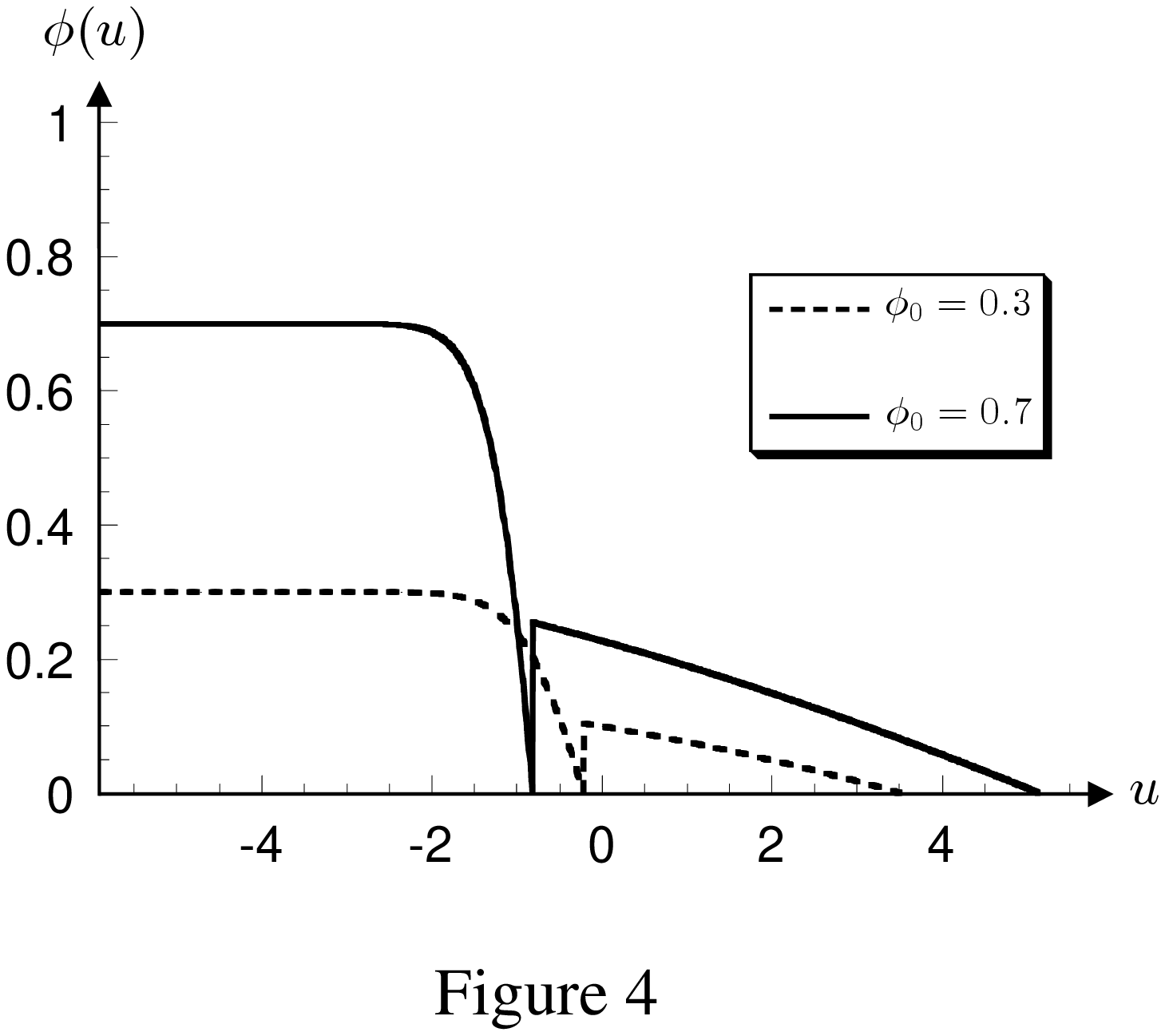} \caption{Numerical profiles for the volume
fraction $\phi(u)$ of the A species (with $u=x/2 \sqrt{\Ds t}$).
The chain length and Flory parameter are $N=10^3$ and
$\chiab=0.1$, yielding $\epsilon= 5 \cdot 10^{-3}$. The solution
$\phi(u)$ has been represented for two values of the initial
volume fraction of the A species in the A-A* blend: $\phi_0=0.3$
and $\phi_0=0.7$.} \label{profilsnumeriques}
\end{figure}

As an illustration of the generic sketch of the diffusion profile
described above, we have plotted in Figure~\ref{profilsnumeriques}
two actual profiles, computed numerically for different values of
$\phi_0$ (and same $\epsilon$), which do show that, within some
rescaling, the generic shape of Figure~3 is retained in both
cases.

\begin{figure}
\centering
\includegraphics*[scale=.7, clip=true, bb=3.2cm 8.2cm 18.5cm
18cm]{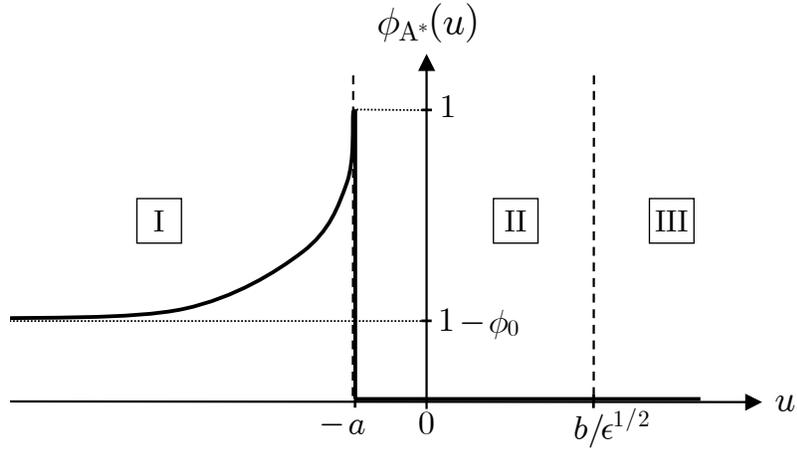} \caption{Sketch of the generic self-similar
diffusion profile for the A* species: the volume fraction
$\phi_\text{A*}$ has been drawn versus the reduced variable $u=x/2
\sqrt{\Ds t}$. In region~I, $\phi_\text{A*}=1-\phi$ and is thus
given by a complementary error function. In the other regions, A*
is absent.} \label{profastar}
\end{figure}

\begin{figure}
\centering
\includegraphics*[scale=.7, clip=true, bb=3.2cm 8.2cm 18.5cm
18cm]{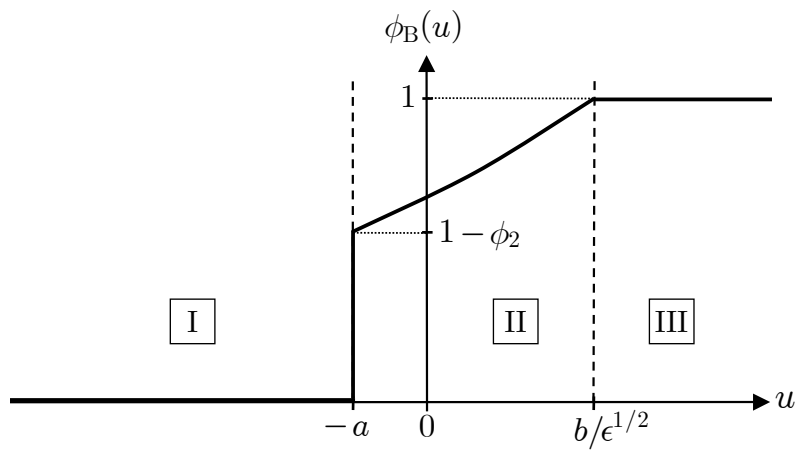} \caption{Sketch of the generic self-similar
diffusion profile for the B species: the volume fraction
$\phi_\text{B}$ has been drawn versus the reduced variable $u=x/2
\sqrt{\Ds t}$. In region~I, B is absent; in region~II, the profile
of B is the complement of A ($\phi_\text{B}+\phi=1$) and is
quasi-linear; in region~III, B is alone.}
\label{profb}
\end{figure}

For completeness, we also give the corresponding generic
self-similar profiles for the other mixture components in
Figures~\ref{profastar} and~\ref{profb}.
As reminded in Figure~\ref{profilgenerique}, the diffusion profile
involves several characteristic quantities $a$, $b$, and $\phi_2$,
whose values we now consider in detail.

\begin{figure}
\centering
\includegraphics*[scale=.5, clip=true, bb=3.2cm 5.3cm 19cm
17.5cm]{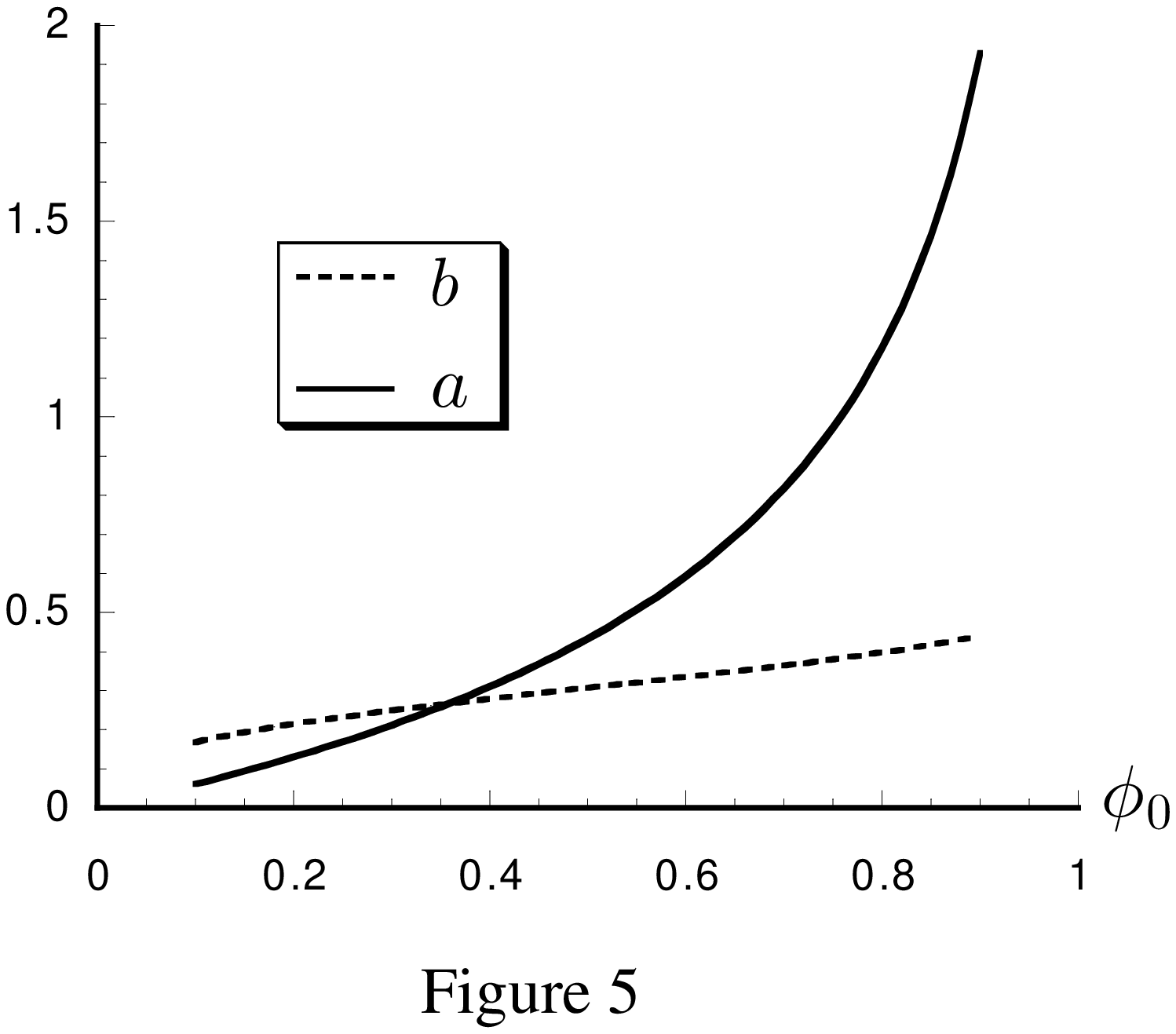} \caption{Values of the numerical factors $a$
and $b$, which characterize the two moving boundaries $\xI(t)=-
2a\,(\Ds t)^{1/2}$ and $\xII(t)=2b\,(\Df t)^{1/2}$, plotted versus
the initial volume fraction $\phi_0$. ($\phi_0$ is ranging from
0.1 to 0.9 and $\epsilon$ is fixed at $5\cdot 10^{-3}$.)}
\label{aandb}
\end{figure}

\begin{figure}
\centering
\includegraphics*[scale=.5, clip=true, bb=3.3cm 5cm 19.3cm
18cm]{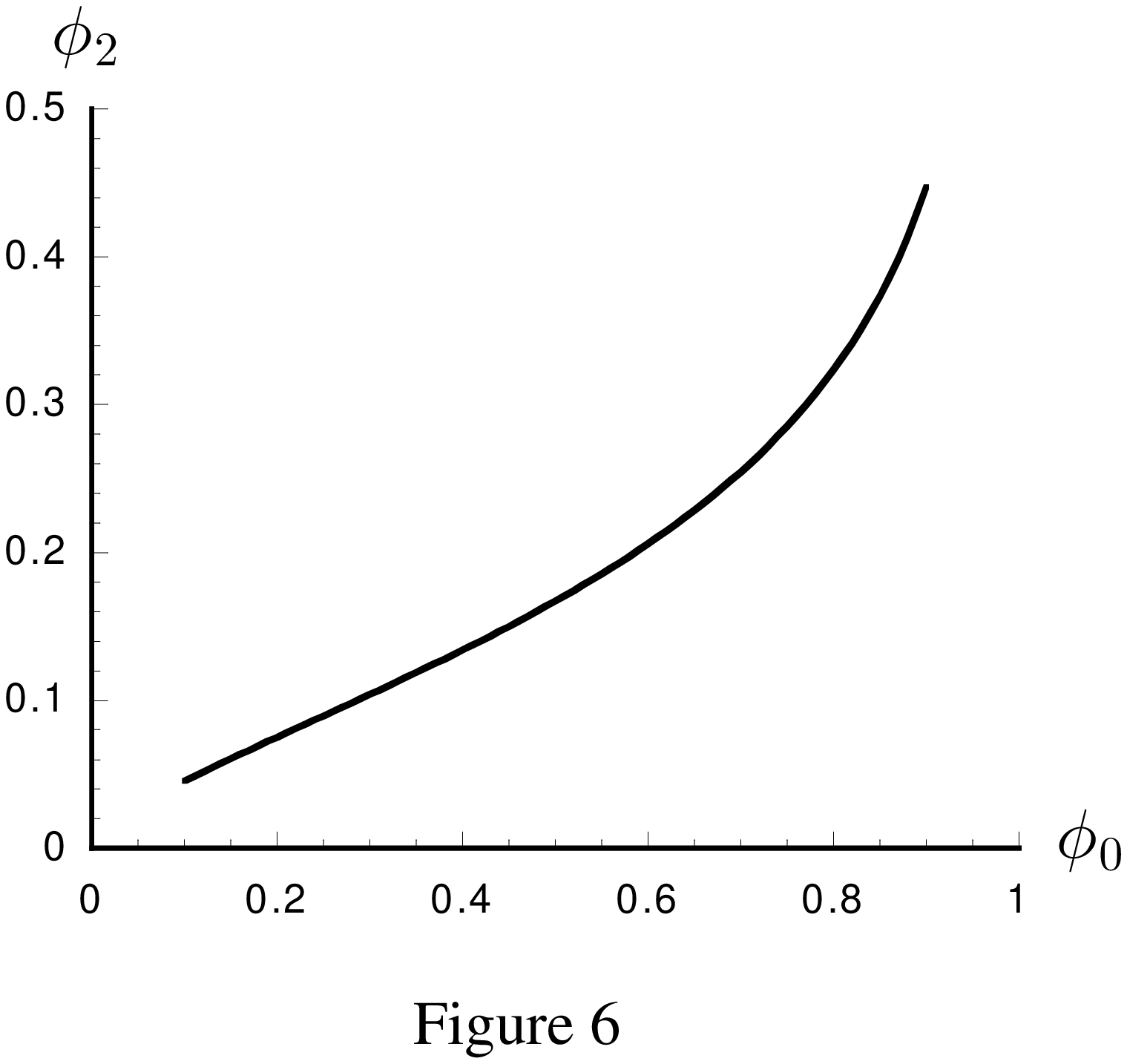} \caption{Plot of the quantity $\phi_2$, which
corresponds to the amplitude of the concentration jump at the
border between region~I and~II, versus the parameter $\phi_0$.
($\phi_0$ is ranging from 0.1 to 0.9 and $\epsilon$ is fixed at
$5\cdot 10^{-3}$.)} \label{phi2}
\end{figure}

In Figures~\ref{aandb} and~\ref{phi2}, we plot the variations of
these quantities as functions of our parameter $\phi_0$, i.e.,
$a(\phi_0)$, $b(\phi_0)$, and $\phi_2(\phi_0)$, at fixed
$\epsilon$ (i.e., we work at given system chemistry, and change
the initial composition of the system). It is observed that these
quantities show a steady increase with $\phi_0$. Results are
presented for the range $0.1\lesssim \phi_0 \lesssim 0.9$, which
should cover most practical cases; see note\cite{limits} about
situations where $\phi_0\to 0$ or $\phi_0\to 1$.

It is also possible to give rough analytical estimates of the
dependence of $a$, $b$, and $\phi_2$ with respect to our other
main parameter, $\epsilon$. The quantity $a$ is given by the
solution of eq.~\ref{equationsura}, which involves only $\phi_0$;
thus we see that $a$ has no functional dependence on $\epsilon$
(at least in the regime $\epsilon \ll 1$ considered in this
article). Let us now estimate the dependencies of $\phi_2$ and $b$
on $\epsilon$: for that purpose, we will consider, in first
approximation, that the concentration profile is linear in
region~II (see above). Writing as previously the solution $\phi(x,
t)=g(v)$ in terms of the reduced variable $v=x/2 (\Df t)^{1/2}$,
the (``constant'') value of the slope $\dd g/\dd v$ is easily
estimated: since $g(v)$ goes from $\phi_2$ at the left boundary of
region~II ($v=-a\sqrt\epsilon$) to zero at the right boundary
($v=b$), we have $\dd g/\dd v \simeq
-\phi_2/(b+a\sqrt\epsilon)\simeq-\phi_2/b$. We then use this
expression of the slope into eq~\ref{conditionssurg}-(i) and
eq~\ref{conditionsupplementaire}, and solve this set of two
equations for $b$ and $\phi_2$. This brings the following
estimates:
    \beq
    \label{scalingbetphi2}
    b \simeq \left(\frac{a^2}{4} \epsilon\right)^{1/6}, \qquad \phi_2 \simeq (2 a^2 \epsilon)^{1/3}
    \eeq
In terms of scaling laws with respect to $\epsilon$, we thus have
    \beq
    \label{approxbetphi2}
    b \sim \epsilon^{1/6}, \qquad \phi_2 \sim \epsilon^{1/3},
    \qquad a\sim\epsilon^0
    \eeq
(where the last equation is meant  to recall that $a$ has no
dependence on $\epsilon$).

Finally, we conclude the presentation of our results, by
emphasizing one unusual and notable feature of our solution: as
time goes on, region~II, where A and B mix, grows in a very
asymmetric fashion, relatively to the position of the initial
interface ($x=0$ at $t=0$); this is due to the fact that its two
boundaries are of very different nature, with the right boundary
$x=\xII$ having a much faster diffusive motion than the left
boundary $x=-\xI$ (because $\Df\gg\Ds$). The ratio of these two
boundaries' position is in fact independent of time and can be
estimated as
    \beq
    \frac{\xI(t)}{\xII(t)} = \frac{a}{b}
    \left(\frac{\Ds}{\Df}\right)^{1/2}\sim\epsilon^{-1/6}\epsilon^{1/2}\sim\epsilon^{1/3}
    \eeq
The sluggishness of the motion of $\xI$ relatively to $\xII$ thus
becomes more pronounced as $\epsilon$ becomes smaller, i.e., if
the attraction between A and B is made stronger or if the polymer
chains length is increased (see eq~\ref{defepsilon}).

We have now completed the solution of the dynamics of our
three-species system, and have presented the results that were
obtained.
\section{Concluding remarks}
\label{discussion}

In this article, we studied the evolution of the interface between
two polymer melts, in a specific case where three species of
strongly contrasting chemical affinities were involved. It was
found that, due to the simultaneous presence of entropy and
enthalpy-driven diffusion processes, the dynamics at the interface
is unusual (with a very asymmetric growth of the mixing layer),
and that moreover, the interface shows a peculiar spatial
structuration in three distinct layers of different chemical
compositions.

Let us now close with some remarks. Obtaining a triad of polymers
with chemical properties such as specified in
eqs~\ref{defchiab}-\ref{defchiaastar} is certainly a difficult
task. One potential way could be the following: (i)~having A and B
strongly attracted to each other through the existence of hydrogen
bonds, (ii)~having A* a very slightly modified variant of A, where
hydrogen bonding to B becomes unavailable; then between A and B,
van der Waals attractions will dominate, leading to repulsion,
while hopefully A and A* will remain similar enough and roughly
indifferent to each other. Another, powerful way of achieving such
a system may also be through the use of \emph{copolymers}.

The specific case which has been described and studied in this
article is, as has been argued in the Introduction, only one
limiting situation in a vast range of multiple species interfaces
in polymer systems, and has been selected for the strong opposing
tendencies in the chemical affinities of the involved polymers.
Our hope is however to demonstrate on this example the richness of
multiple-species interfaces, where many other such ``selected
cases'' would be worth investigating and studying for their
peculiar structure and dynamics.

A related, though different, problem to that of two
multi-component melts facing each other, is the situation of a
melt facing a \emph{loosely crosslinked gel} (this perspective was
suggested to us by one of the referees). In the example studied in
this article, an A+A* blend would be brought into contact with a
gel of B. The subsequent evolution might develop interesting
features: the motion of the $\xII(t)$ front (as defined on Fig.~2)
would describe the imbibition of the B gel by the A species, while
the displacement of the $\xI(t)$ border would be dictated by the
swelling of the gel---this would imply the inclusion of new
elastic terms in the model's equations. We note however that, in
practice, the dynamics of polymer diffusion into a network can
prove very subtle, with a predominant role of preexisting
heterogeneities (e.g., in crosslink density) within the
network.\cite{RussNetwork}

From the application point of view, the ability to spontaneously
form (at least in some cases) a \emph{multi-layered interface}
between two polymer pieces, might lead to interesting perspectives
in the design of multilayer polymer
films\cite{multilayer}---~which are widely used for product
packaging, for instance. Whereas conventional technology includes
processes like coating, lamination, or coextrusion to form
multilayered films, here the layering appears by means of a purely
diffusional process. The chemical composition and the spatial
organization of the layering could be controlled through an
appropriate choice of chemical affinities between the different
species in interaction, and once the desired result is attained
(e.g., prescribed layer thicknesses), the system could be quenched
to freeze the diffusion process.

\section*{Acknowledgements}
We are very grateful to F.~Brochard-Wyart, F.~Chevy, C.~Creton and
M.~Winnik for useful discussions. We also wish to thank our
anonymous reviewers for several deep comments which led to
improvements in the article.
\section*{Appendix~A. Transient behavior at the onset of contact}
In this Appendix, we discuss the transient behavior taking place
in the first instants after the contact between the initial melts
is established, and how it leads to the layered configuration
depicted on Figure~2 of the main text.

Let us consider the situation immediately after the onset of
contact (see Fig.~1), from the point of view of species B: due to
thermal agitation, a few B molecules start to extend chain
portions into the A+A* blend, of respective volume fractions
$\phi_\text{A}$ and $\phi_\text{A*}$. These first B chain portions
will then experience an A+A* environment with an average,
``effective'' Flory parameter
    \beq
    \label{chiBaverage}
    \overline\chi_\text{B}\simeq\phi_\text{A} \chiab +
    \phi_\text{A*}\chiastarb
    \eeq
(in a mean-field picture). Given the opposite signs of $\chiab$
and $\chiastarb$, this effective $\overline\chi_\text{B}$ may be
positive or negative depending on the composition of the A+A*
blend. Defining
    \beq
    \phic=|\chiastarb|/(\chiab+|\chiastarb|)
    \eeq
we have two cases depending on the initial ($t=0$) content
$\phi_\text{A}=\phi_0$ in A of the A+A* blend: either
$\phi_0<\phic$ or $\phi_0>\phic$. We discuss below the two
scenarios that ensue for the transient evolution of the system at
short times.

If the initial situation is such that $\phi_0<\phic$, we see from
eq~\ref{chiBaverage} that $\overline\chi_\text{B}>0$, which means
that the environment seen by the B chains exploring the initial
A+A* blend is enthalpically unfavorable. Therefore the triple
region A+A*+B created by this diffusion remains of very small
extension and corresponds to our region~IV in Fig.~2. On the other
hand, the diffusion of species A into the B melt is very favorable
($\chiab<0$), and thus A molecules do at the same time migrate
from the initial A+A* blend through the triple region~IV, and
create a mixing layer A+B as seen in Fig.~2. At that point in
time, the layered structure of the interface is established and
the subsequent dynamics of the system crosses over to that
described in the main text (see Sec.~\ref{qualitativeapproach}).

Alternatively, if the initial situation of the system is such that
$\phi_0>\phic$, eq~\ref{chiBaverage} tells us  that
$\overline\chi_\text{B}<0$ and in that case, the environment seen
by the B molecules initially in contact with the A+A* blend is
\emph{favorable}. Therefore, it is expected that after the onset
of contact, B molecules will start diffusing into the A+A* blend
and thus a \emph{growing} region where the three species A, A* and
B overlap will form. This may at first seem at odds with the
layered configuration of Fig.~2 and especially with the assumption
of a \emph{thin} triple region invoked in the main text. However,
if one considers the point of view of species A and A*, simple
energetic arguments make it clear that the initial growth of the
triple region will be short-lived: firstly, the triple region
makes an unfavorable environment for A* (the local effective Flory
parameter for A*, as an average of $\chiaastar\simeq 0$ and
$\chiastarb>0$, is always positive) and thus A* will demix and
migrate back to the A+A* melt on the left; secondly, rather than
staying within the triple region where part of their contacts are
made with A* molecules (with no enthalpy gain), A chains will seek
to augment favorable $\chiab<0$ contacts and will thus preferably
migrate into the pure B region on the right of the triple
region---thereby creating the A+B region described in the main
text as region~II. As both these processes occur at the expense of
the triple region, the size of the latter can only remain modest.
Qualitatively at least, it thus becomes apparent, that here again,
the system rapidly evolves to the configuration of Fig.~2. A more
precise description of the transient dynamics just sketched would
nevertheless be desirable and remains ahead of us.

(Note: in the limit where the initial A+A* blend is in fact almost
purely composed of A, i.e., $\phi_0$ very close to unity, we
expect some of the above arguments to break down; the system will
not form layers at the interface, but should simply evolve as in
the known situation of a pure melt of A in contact with a pure
melt of B.\cite{BJL})
\section*{Appendix~B. Derivation of governing equations}
In this Appendix, we give a detailed derivation of some of the
governing equations presented in section~\ref{governing}.
\subsubsection*{Equation
of motion for \boldmath{$\xII(t)$}}
Equation~\ref{equationxIIpoint} is obtained as follows. By
definition, $\xII(t)$ is the point such that, at any time $t$,
$\phi(\xII(t),t)=0$. Since $\phi$ is a constant over time at that
point, we also have that the total time derivative at that
(moving) point, $\dd\phi/\dd t|_{x=\xII}$, must be zero:
$$\left.\frac{\dd\phi}{\dd
t}\right|_{x=\xII}=\dot\phi|_{x=\xII}+\dotxII.\nabla\phi|_{x=\xII}=0$$
Furthermore, using the diffusion equation~\ref{diffeqII} and
$\phi|_{x=\xII}=0$, another expression for $\dot\phi$ is easily
obtained: $$\dot\phi|_{x=\xII}= \Df(\nabla \phi|_{x=\xII})^2$$
Substitution of the latter equation into the former yields
equation~\ref{equationxIIpoint}.
\subsubsection*{Conservation of the B species}
We now show how relation~\ref{conservationBlocale} results from
the conservation of the B species.

As a little reflection from the consideration of Figures~1 and~2
can make clear, the conservation of the B species reads as
follows:
    \begin{equation}
    \label{conservationBintegree1}
    \int_{-\xI(t)}^{\xII(t)}\phi_\text{B}\,\dd x=\xII(t)\cdot 1
    \end{equation}
This expression simply states that, when considering the situation
at a certain time $t>0$, all the initial B material which was
comprised between $x=0$ and the present position $x=\xII(t)$ (a
quantity equal to $\xII\cdot 1$ since the initial volume fraction
of B was one) has been redistributed by virtue of the
interdiffusion process all over region~II (hence the term
$\int_{-\xI}^{\xII}\phi_\text{B}\,\dd x$ on the left-hand side).
Using $\phi_B=1-\phi$, one then obtains
    \begin{equation}
    \label{conservationBintegree2}
    \int_{-\xI(t)}^{\xII(t)} \phi \,\dd x = \xI(t)
    \end{equation}
Differentiating eq~\ref{conservationBintegree2} once with respect
to time, using eq~\ref{diffeqII} and $\phi(\xII)\equiv 0$, one
obtains a local version of this integral equation: $
\dotxI=-\phi_2\cdot\Df\nabla\phi|_{x=-\xI^+}$, which is exactly
equation~\ref{conservationBlocale}.
\subsubsection*{Conservation of the A species}
We now derive equation~\ref{conservationAlocale} from the
conservation of species A.  The conservation equation for A has
the form
    \beq
    \label{conservationAintegree}
    \int_{-\xI(t)}^{\xII(t)} \phi\,\dd x =
    \phi_0\xI(t)+\int_{-\infty}^{-\xII(t)}(\phi_0-\phi)\,\dd x
    \eeq
This expression has the following meaning: at time $t$, the A
material enclosed in region~II ($\int_{x=-\xI}^{x=\xII} \phi\,\dd
x$) comes from the material which was initially present between
$x=-\xI$ and $x=0$ at volume fraction $\phi_0$, plus the amount
that has diffused from region~I
($\int_{-\infty}^{-\xII}(\phi_0-\phi)\,\dd x$). Using
eq~\ref{conservationBintegree2} into
eq~\ref{conservationAintegree}, then differentiating with time and
substituting with the diffusion eq~\ref{diffeqI}, one finally
obtains the following local version for the conservation of the A
species: $\dotxI=-(1-\phi_1)^{-1}\cdot\Ds\nabla\phi|_{x=-\xI^-}$,
which is precisely equation~\ref{conservationAlocale}.
\subsubsection*{Relation between \boldmath{$\phi_1$ and $\phi_2$}}
We now explain how one obtains equation~\ref{equationphi1phi2}
from~\ref{egalitepotentiel}.

Using classical
formulae~\cite{PGGScalingConcepts,JonesRichardsPolymerInterfaces,PetitDoi}
to compute the chemical potential of A, one obtains the expression
    \beq
    \label{potentielchimique}
    \muA=kT\bigl[\log\phi+\chi N(1-\phi)^2\bigr]
    \eeq
This expression for \muA holds for blends made of \emph{two}
polymer species, characterized by a Flory parameter $\chi$ (and
under the assumption that these species have the same chain length
$N$). Thus, eq~\ref{potentielchimique} should not be used to
compute \muA within region~IV, where three component coexist; but
it can indeed be used at the border with region~I on the left, and
at the border with region~II on the right---there one of the
components drops to zero concentration, thereby leaving us with
two species only (see also note~\cite{notepotentiel}).

On the left border of region~IV ($x=-\xI^-$), we only have the two
species A and A* (since $\phi_\text{B}\to 0$), and accordingly, we
can apply formula~\ref{potentielchimique} with
$\chi=\chiaastar\simeq0$, which yields
$$\frac{\muA(-\xI^-)}{kT}=\log\phi_1$$ Similarly, on the right border
($x=-\xI^+$), only A and B coexist, and we find
$$\frac{\muA(-\xI^+)}{kT}=\log\phi_2 + \chiab N(1-\phi_2)^2$$ Writing
the equality between the above two expressions of the chemical
potential (as required by eq~\ref{egalitepotentiel}) yields the
governing equation~\ref{equationphi1phi2}.

\begin{thebibliography}{99}
\bibitem{notepotentiel}
If conditions are such that region~IV is so thin as to be
comparable to the radius of gyration of the chains, chemical
potentials cannot be defined within this region (because
non-local, square-gradient terms enter into the system's free
energy). In that case, the chemical potentials appearing in
equality~\ref{egalitepotentiel}, which are calculated ``at the
left and right boundaries'' of region~IV, should naturally be
understood as evaluated slightly away from those boundaries
(typically a few radii of gyration), where their definition makes
sense again.
\bibitem{limits}
Results in the main text are presented for initial fractions
$\phi_0$ of the A species comprised between 0.1 and 0.9 which
cover most of the interesting situations. Outside this range, some
of the assumptions used in the solution of the model do break
down; however, the physical situation is simple since it is
basically getting back to usual two-species configurations (as
described in the overview, section~\ref{reminder}), with no
``interfacial layering'' at all. For the smaller fractions
($\phi_0\to 0$), the system tends to the situation of an (almost)
pure melt of A* facing a B melt (with a sharp A*/B interface in
between). For the higher initial fractions ($\phi_0\to 1$), the
evolution is on the contrary that of an (almost) pure A melt
facing a pure B melt, with enthalpic mixing taking place.
\end{thebibliography}
\end{document}